\documentstyle[sprocl,epsfig]{article}

\def\Pom{{\bf I\!P}}

\def\Journal#1#2#3#4{{#1} {\bf #2}, #3 (#4)}

\def\ZPC{{\em Z. Phys.} C}

\def\be{\begin{equation}}
\def\ee{\end{equation}}
\def\bea{\begin{eqnarray}}
\def\eea{\end{eqnarray}}

\begin{document}

\title{DIFFRACTIVE STRUCTURE FUNCTION AT VERY SMALL $\beta$
AND UNITARITY CORRECTIONS IN THE COLOR DIPOLE REGGE APPROACH}

\author{W. SCH\"AFER}

\address{Institut f\"ur Kernphysik, FZ-J\"ulich, 
\\ D-52425 J\"ulich, Germany\\E-mail: wo.schaefer@fz-juelich.de} 


\maketitle\abstracts{Starting from the phenomenologically successful color dipole
representation of the diffractive structure function,
we identify the dominant contribution to the diffractive
cross section at small beta. The beta dependence is calculated
taking advantage of the Regge factorization properties 
of the hard mechanism which drives the dependence on the
diffractive mass. We impose the unitarity constraints on
the $x_\Pom$-dependence and comment on the screening 
corrections to $F_{2p}$.}

Deep inelastic scattering at large values of the Regge
parameter $1/x = (W^2 + Q^2) / Q^2 \gg 1$ ( with
$W^2 = (p+q)^2$ the $\gamma^*$-target cms-energy squared,
and $Q^2 = -q^2$, the photon virtuality) is conveniently
viewed
in terms of the scattering of the $q\bar{q}$ color--dipoles  
of sizes $\bf{r}$ and $\bf{r'}$ in the photon and
target $t$  respectively. Making use of the color
dipole (CD) factorization the $\gamma^* t$ cross section
takes the form:
\bea
\sigma^{\gamma^* t}(x,Q^2) = \int dzd^2{\bf{r}} dz' d^2{\bf{r'}}
\left|\Psi_{\gamma^*} (z,{\bf{r}}) \right|^2 \left|\Psi_t(z',{\bf{r'}})
\right|^2
\cdot \sigma(x,{\bf{r}},{\bf{r'}}) \, .
\eea
Here $|\Psi_{\gamma^*,t} (z,\bf{r})|^2$ denotes the probability
to find a $q\bar{q}$ dipole of tranverse size and orientation 
${\bf{r}}$  in the $\gamma^*,t$ with a partitioning $z,1-z$
of the $\gamma^*,t$--lightcone momenta between its constituents.
The principal dynamical quantity of the color dipole approach
is the dipole--dipole cross section $\sigma (x, {\bf{r}},{\bf{r'}})$,
and the main feature of the CD-Regge approach is
the expansion of the CD-cross section in terms
of isolated Regge poles:
\bea
\sigma(x,r,r') = \sum_m C_m \sigma_m(r) \sigma_m(r') \, 
\left({x_0\over x}\right)^{\Delta_m} \,
\eea
which uniquely determines the $1/x$--dependence of the dipole--dipole
 cross section (and hence of $\sigma^{\gamma^* t}$) in terms 
of the eigen-cross sections $\sigma_m(r)$. The latter are
solutions to the color dipole equation discussed in \cite{NZZ},
taking the Regge form $\sigma_m(x,r) = \sigma_m(r) (x_0/x)^{\Delta_m}$.
We recall that the $1/x$--dependence of the dipole cross section
takes into account the effect of multigluon states in the
multiperipheral Fock-state wave function of the $\gamma^*$.
For the discussion of the solutions, intercepts, and slopes, see \cite{NZZ},
where a succesful phenomenology is also described.
The color--dipole formulation is by now a standard approach to diffractive
DIS \cite{Diff}, which emerges naturally as the quasielastic scattering of the 
$q\bar{q}$ (small diffractive masses $M$, large $\beta
= Q^2/(Q^2 + M^2)$ and $q \bar{q}g$ (large
$M$, small $\beta$) components of the virtual photon.
While the former can be reinterpreted with some care in 
terms of the valence quark content of the Pomeron, the
latter can be associated with the photoabsorption on a sea--quark
in the Pomeron, which in turn was radiatively generated from
a 'valence--gluon' in the Pomeron.
The corresponding two--gluon wavefunction of the Pomeron
was derived in \cite{NZ94}:
\bea
|\Psi_\Pom(z_g,{\bf{\rho}},x_\Pom )|^2 = {1 \over \sigma^{pp}_{tot}} {1\over z_g}
\left[ {\sigma_{gg} (x_\Pom , \rho) \over \rho^2} \right]^2 \cdot
{\cal{F}}(\rho/R_c ) \, ,
\label{PWF}
\eea
and is the starting point of our generalization of the CD-Regge expansion to
diffractive DIS (DDIS). Notice its explicit dependence on the Regge parameter
$1/x_\Pom = (W^2+Q^2)/(M^2+Q^2)$; furthermore it is nonsingular
at the origin $\rho = 0$, which makes the emerging triple Pomeron coupling
dominated by the nonperturbative scale $R_c \sim 0.25-0.3$ fm. The latter
enters through the infrared cutoff function ${\cal{F}}$ and must be 
viewed as a finite propagation radius of the perturbative gluons.
Two major conclusions follow from eq. (\ref{PWF}): first, the two
gluon wave--function prescribes how the Pomeron enters as a target 
in the CD--Regge expansion, which can now be straightforwardly
applied to the diffractive structure function, alias the $\gamma^* \Pom$
total cross section. This allows us to build up the full $\beta$--dependence
associated with the triple--$\Pom$ regime of DDIS at fixed value of $x_\Pom$.
And second, we make the crucial observation that the strong dependence 
on $x_\Pom$ must be subject to the unitarity constraints for the
$s$--channel partial waves. In view of the parameter $R_c^2/R_p^2 \ll 1$,
the natural quantity is the dipole--quark scattering profile function
($s$--channel partial wave amplitude)
\bea
\Gamma_0 (x_\Pom, \rho, {\bf{b}}) = {1\over 3} \cdot {\sigma_{gg} (x_\Pom, \rho)
\over 4 \pi B(x_\Pom)} \cdot \exp\left[{-b^2 \over 2B(x_\Pom)} \right] \, .
\eea
Here $\sigma_gg$ is the cross section for scattering of a $gg$--dipole on a nucleon,
and the Regge slope $B(x_\Pom) = 1/3 R_c^2 + 2 \alpha'_\Pom \log(x_0/x_\Pom)$,
$\alpha'_\Pom \sim 0.07 GeV^-2$. The essence of $s$--channel unitarity
is the bound $\Gamma (x_\Pom,\rho, {\bf{b}}=0) \leq 1$ on the relevant 
partial waves which we achive by means of plugging the bare $\Gamma_0$
into an eikonal formula yielding the unitarized cross section $\sigma^U(x_\Pom, \rho)$,
and hence the unitarity corrected $|\Psi^U_\Pom (z_g,\rho,x_\Pom)|^2$.
Our procedure sums the multipomeron exchanges which couple to the lowest 
ladder cell in the cut Pomeron, and hence allow the full $\beta$--span
to evolve from small size $1/Q$ at the $\gamma^*$ vertex to large size 
$R_c$ at the lower end \cite{NZ94}. 
In the figure we show a sample prediction \cite{NSZ} of
the diffractive structure function $F_2^{D(3)} \propto \int dp_\perp^2 
x_\Pom d\sigma(\gamma^* p \to pX)/dx_\Pom dp_\perp^2 $ at $x_\Pom = 0.03$.

\begin{figure}[h]
\epsfxsize=0.6\hsize
\epsfbox{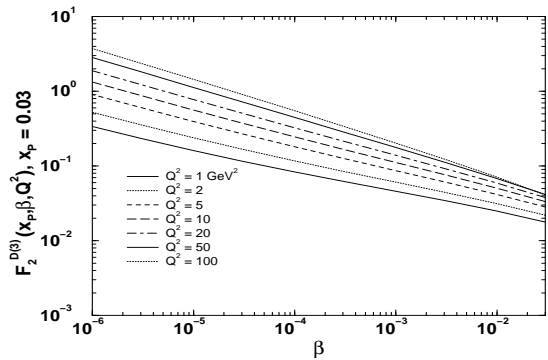}
\caption{The diffractive structure function 
$F_2^{D(3)} (x_\Pom,\beta,Q^2)$ as a function of $\beta$ at $x_\Pom = 0.03$
for several values of $Q^2$.}
 \label{fig1}
\end{figure}

The opening of diffractive channels comes along with the unitarity (screening,
shadowing, absorption)
correction to the proton structure function, for which a 
standard AGK \cite{AGK} estimate can be obtained by integrating 
the diffractive cross section over masses. Our calculation gives
$R_{sh} = \Delta^{sh}F_{2p}/F_{2p} \sim 0.3$ at $x\sim 10^{-4}, Q^2 = 1 GeV^2$
growing to $R_{sh} \sim 0.35$ at  $x\sim 10^{-6}$. 
Hence we see no sign of reaching the black disc limit $R_{sh} = 0.5$ 
any soon. However we stress that the found unitarity corrections
are large enough to provide a substantial systematic uncertainty
for all two--parton ladder based DIS--phenomenology and may render some
of its underlying assumptions unwarranted.

\end{document}